\documentclass[11pt,twoside]{article}
\usepackage{asp2004}
\usepackage{psfig}
\usepackage{epsf}
\usepackage{graphics}
\usepackage{lscape}
\def\edcomment#1{\iffalse\marginpar{\raggedright\sl#1\/}\else\relax\fi}
\reversemarginpar

\begin{document}
\title{The Cluster M/L Ratio and the Value of \boldmath $\Omega_m$}
\author{H. Andernach}
\affil{Departamento de Astronom\'{\i}a, Universidad de Guanajuato, \\
Apartado Postal 144, 36000 Guanajuato, Gto, Mexico}
\author{M. Plionis\,$^{1,2}$, O. L\'opez-Cruz\,$^{2}$, E. Tago\,$^{3}$, S. Basilakos\,$^{1}$}
\affil{$^{1}$ Institute of Astronomy \& Astrophysics, I. Metaxa \& V. Pavlou, \\
\hspace*{2mm} Palaia Penteli, 15236 Athens, Greece}
\affil{$^{2}$ Instituto Nacional de Astrof\'\i sica, Optica y Electr\'onica (INAOE),
\hspace*{2mm} Apartado Postal 51 y 216, 72000 Puebla, Pue., Mexico}
\affil{$^{3}$ Tartu Observatory, EE--2444 T\~oravere, Estonia}

% \author{M. Plionis}
% \affil{Inst. of Astronomy \& Astrophysics, I. Metaxa \& V. Pavlou, 
% Palaia Penteli, 15236 Athens, Greece;
% INAOE, Puebla, Pue., M\'exico}
% \author{O. L\'opez-Cruz}
% \affil{Instituto Nacional de Astrof\'\i sica, Optica y Electr\'onica (INAOE), 
% AP 51 y 216, 72000 Puebla, Pue., Mexico}
% \author{Erik Tago}
% \affil{Tartu Observatory, EE--2444 T\~oravere, Estonia}
% \author{S. Basilakos}
% \affil{Inst. of Astronomy \& Astrophysics, I. Metaxa \& V. Pavlou,
% Palaia Penteli, 15236 Athens, Greece}

\begin{abstract}
From an up-to-date compilation of ACO cluster redshifts and velocity dispersions 
we extract a homogeneous sample of 459 clusters with robust velocity dispersion.
Using the virial theorem to estimate cluster masses, and a
correlation between Abell galaxy counts and $R$-band luminosity,
we find a median $M/L_{R} \simeq 174\,h_{50}\;M_{\odot}/L_{\odot}$.
However, if we correct the virial masses according to the 
X-ray vs.\ virial mass correlation, as derived from our analysis, 
we obtain $M/L_{R} \simeq 100 \; h_{50} \; M_{\odot}/L_{\odot}.$
These two values of $M/L$ imply a mass content of the Universe of
$\Omega_{\rm m} \simeq 0.26$ and 0.15, respectively.
\end{abstract}

% \section{Introduction}
% Different cluster studies yielded a wide range of values for the
% mass-to-light ratio in clusters: $M/L \sim 100 - 500 \; h_{50} \;
% M_{\odot}/L_{\odot}$, the largest sample (89 clusters, Girardi et al.\ 2000) 
% gave $M/L_{B} \sim 130 \; h_{50} \; M_{\odot}/L_{\odot}$, which
% then translates to $\Omega_{\rm m} \simeq 0.19$.
% Although this value of $\Omega_{m}$ points towards a low-$\Omega_{\rm
% m}$ Universe, it is still $\sim$ 40\% less than the preferred value of the
% standard Cosmological paradigm.
% Here we use a subsample of 459 clusters with very robust values of
% the velocity dispersion, $\sigma_V$ to determine cluster masses
% from the virial theorem.  Then we determine cluster luminosities based
% on a sample of 46 X-ray selected Abell clusters
% with excellent photometry.  From this we derive the individual 
% cluster $M/L$, the distribution of which is used to derive 
% $\Omega_{\rm m}$. We use $H_{\circ}=50\;h_{50}$km\,s$^{-1}$\,Mpc$^{-1}$,
% $\Omega_{\rm m}=0.3$, and $\Omega_{\Lambda}=0.7$ to calculate distances.
% See Plionis et al.\ 2004 for a more detailed version of the present work.

\section{Cluster Velocity Dispersions}

We made use of the July 2003 version of a compilation (Andernach et al.\ 2004) 
of published redshifts and velocity dispersions ($\sigma_{v}$) of clusters from 
the Abell, Corwin \& Olowin (1989, ACO) catalogue. This version gives $\sigma_{v}$ for 1595 
Abell/ACO cluster components of 1240 unique Abell/ACO clusters
(1002 A-clusters and 237 supplementary S-clusters),
often with two or more components at different redshift $z$ 
(subclusters or line-of-sight components).  For the latter
we take as the main cluster that component which minimizes
$|z_{i}-z_{phot}| \times n_{z,t}/n_{z,i}$
(where $z_i = \overline{z}$ of $i^{th}$ component, $z_{phot} =$ photometric cluster redshift, 
$n_{z,t} =$ total number of galaxy redshifts for that cluster,
$n_{z,i} =$ number of redshifts in the $i^{th}$ component). 
We define a corrected Abell richness count as ~$N_{Ab,c}=N_{Ab} \times n_{z,i}/n_{z,t}$,
and keep only clusters with $N_{Ab,c}\ge30$ (i.e.\ Abell richness $R\ge0$).
We use $H_{\circ}=50\;h_{50}$km\,s$^{-1}$\,Mpc$^{-1}$,
$\Omega_{\rm m}=0.3$, and $\Omega_{\Lambda}=0.7$ to calculate distances.
%  or Abell richness class $R\ge0$.

To limit ourselves to robust values of $\sigma_v$,
we perform tests (cf.\ Plionis et al.\ 2004 for details) and find that only
for those 60\% of clusters with $n_z\ge\,12$ may we consider $\sigma_v$ as 
reliable. This also lowers the mean $z$ of our sample, implying that 
$z$ decreases with $n_z$, i.e.\ observers tend to study more 
thoroughly nearby clusters, while
the well-sampled, distant clusters are likely the richest ones.
The observed $n_z\,-\,N_{Ab}$ correlation in turn suggests a
correlation between $N_{Ab}$ and $z$. Thus, to further guarantee
that our sample is a fair representation of the universal mix of 
poor and rich clusters, we perform further tests and restrict our sample 
to $z\le 0.13$.  This leaves a ``clean'' sample of velocity dispersions for
459 $R\ge\,0$ A-clusters (excluding any secondary components) with  
$n_z\ge 12$ and $z\le 0.13$. The median $\sigma_{v}$
and its 67\% and 33\% quantiles are~
$\sigma_v \approx 704^{+94}_{-119} \; {\rm km\,s}^{-1}.$
The $\sigma_{v} - N_{Ab,c}$ relation is
$\sigma_v/{\rm km\,s}^{-1}  \simeq 2.32 (\pm 0.42) N_{Ab} + 562 (\pm 27).$

\section{Estimating Cluster Masses: Virial vs.\ X-ray Masses}

We derive the cluster mass from the virial theorem assuming 
clusters to be in dynamical equilibrium. If galaxies trace 
the underlying matter distribution we can write (The \& White 1986):
$M_{\rm c}= (3\pi/2) (\sigma_v^2~r_{\rm vir,p}/G) (1-\Delta),$
where $\Delta$ allows for the cluster not being fully enclosed within 
the sampling radius, and depends on the velocity anisotropy of
galaxy orbits (Girardi et al.\ 1998; 2000).
The projected virial radius, $r_{\rm vir,p}$, depends on
the galaxy distribution of a particular cluster. Girardi et
al.\ (1998) express $r_{\rm vir,p}$ as function of
the virial radius, $r_{\rm vir}$, and the cluster core radius, 
$r_c \simeq 0.05~r_{\rm vir}$ (Katgert, Biviano \& Mazure 2004).
With \,$r_{\rm vir}\,=\,4\;\sigma_v/(1000\;{\rm km\,s^{-1}})\;h_{50}^{-1}\;{\rm Mpc}$
\,(Girardi et al.\ 1998) we arrive to
$M_{\rm c} \approx 1.094 \times 10^{15}\,(\sigma_v/1000)^{2} r_{\rm vir,p}\,(1-\Delta)\,M_{\odot}.$
In the absence of data for individual clusters we take 
the median $\Delta$=0.19 from Girardi et al.\,(1998). 

For the cluster X-ray masses we use data from Jones \& Forman (1999, JF99) 
who determined gas masses and temperatures within 1 $h_{50}^{-1}$\,Mpc for 
$\sim$100 clusters in our sample. As our virial mass is that within $r_{\rm vir}$
we convert this mass to that within the common radius 
of 1~$h_{50}^{-1}$ Mpc to be comparable with the X-ray masses.
We use a King profile for the cluster galaxy radial density distribution.
We find a good correlation of virial and X-ray masses 
albeit with virial masses $M_{\rm c}$ being systematically larger than 
X-ray masses $M_{\rm X}$.  One reason may be an underestimate 
of~$\Delta$, while another could be that the virial assumption is not valid 
(despite our careful selection of only clusters with either unimodal 
velocity distributions or the main component of multi-component clusters).
Assuming that X-ray masses are closer to the true cluster masses, 
we corrected our virial masses from $M_{\rm c}$ to $M_{\rm vir}$ using 
the coefficients of the $M_X$--$M_c$ fit.

\section{Cluster luminosities}
To determine the correlation between luminosity and cluster richness, 
we estimate the cluster luminosities using the LOCOS sample
% (Low-redshift Cluster Optical Survey, 
(L\'opez-Cruz 1997; 2001) of 46 local X-ray selected Abell clusters.
L\'opez-Cruz (1997) generated luminosity functions (LF) for each cluster,
using color information to reduce background contamination, and fitted LFs
with one or two Schechter functions.  We determined the LOCOS cluster
$R$-band luminosity, $L_{R}$, by integrating the fitted LFs.

As the rest-frame area covered by each $23^{'}\,\times\,23^{'}$ cluster frame 
is different for each cluster, we put the cluster luminosities on a common 
physical scale of one Abell radius $R_{A}=\;3\;h^{-1}_{50}$\,Mpc. 
Instead of $r_{\rm vir,p}$ we use $D_{\rm L} \tan(\theta)/2$, where
$D_{\rm L}$ is the luminosity distance %for a cosmological model with
$\Omega_{\rm m}=0.3$ and $\Omega_{\Lambda}=0.7$ and $\theta=23^{'}$.
As we have no measure of the optical luminosity for the majority of our 459 clusters, 
we estimate their luminosity from the $L_{R} - N_{Ab}$ correlation of the
LOCOS sample, valid for $N_{Ab}>30$:
$L_{R}/10^{12} L_{\odot} \simeq 0.034 (\pm 0.006) N_{Ab} +  3.92 (\pm 0.56).$
This procedure gives a good approxiamtion to the overall distribution of luminosities.

\section{The Cluster M/L Ratio and the Value of \boldmath $\Omega_{m}$}

In order to derive consistent values of the cluster $M/L$ we reduce the optical 
luminosities to the same radius (1 $h^{-1}_{50}$ Mpc) at which the cluster masses, 
corrected for the $M_{\rm vir} - M_X$ relation, were estimated.
The distribution of the resulting 459 $M/L$ values is slightly 
non-Gaussian, and the median with 67\% and 33\% quantiles is~
$M_{X}/L_{R} \simeq 100^{+13}_{-14} \; h_{50} \; M_{\odot}/L_{\odot}.$
We test the robustness of our procedure by estimating $M/L$ for two
different cluster subsamples for which we have individually-determined
X-ray masses and/or luminosities: \linebreak[4]
(a) for the 34 clusters the
LOCOS and JF99 samples have in common, we use directly the individual 
$R$-band luminosities and X-ray masses, and find (within $1 \; h_{50}^{-1}$ Mpc):
$(M/L_R)_{\rm LOCOS}\simeq  107^{+16}_{-18}\,h_{50}\,M_{\odot}/L_{\odot};$
~(b) for 146 clusters of JF99, using X-ray masses from JF99 and
optical luminosities from our $L_{R}-N_{Ab}$ relation, we obtain~
$(M/L_R)_{\rm JF99}\simeq  104^{+20}_{-22}\,h_{50}\,M_{\odot}/L_{\odot}.$

To compare our results with those of other studies we
translate them to the $R$ band using typical colors of early-type
galaxies: $B-R\simeq 1.45$, $B-V \simeq 0.9$ (Girardi et al.\ 2002) 
and $(B-R)_{\odot}=1.2$, $(B-V)_{\odot}=0.64$, 
from which we obtain
$L_R/L_{R,\odot}=1.26\,L_B/L_{B,\odot}$  ~and~
$L_B/L_{B,\odot}=0.79\,L_V/L_{V,\odot}.$ From
$B-B_{j} \simeq 0.252$ and $(B-B_{j})_{\odot}=0.15$ 
(Eke et al.\ 2004) we obtain
$L_R/L_{R,\odot}=1.14\,L_{B_{j}}/L_{B_{j},\odot}.$
Based on a study of eight clusters with well-determined {\sc ASCA} X-ray masses,
Hradecky et al.\ (2000) find $M/L_{R} \simeq 127\,h_{50}\;M_{\odot}/L_{\odot},$
while Girardi et al.\ (2000), based on virial masses, obtained
$M/L_{R} \simeq 110 \; h_{50} \; M_{\odot}/L_{\odot},$ 
both in good agreement with our results.

Using the virial masses directly, not correcting for the $M_{\rm vir}\;-\;M_X$ relation,
we obtain $M_{\rm vir}/L_{R} \simeq 176^{+46}_{-54}\,h_{50}\,M_{\odot}/L_{\odot}$,
also in excellent agreement with recent studies based either on virial or 
weak-lensing cluster mass estimations. 
For example, Eke et al.\ (2004), using rich galaxy groups in 2dFGRS, 
find $(M/L)_{R}\;\simeq 192\;h_{50}\;M_{\odot}/L_{\odot}$, 
while Sanderson \& Ponman (2003), based on 66 virialized systems, find
$(M/L)_{R}\;\sim 150\;h_{50}\;M_{\odot}/L_{\odot}$. 
Finally, Hoekstra et al.\ (2002) using an HST weak-lensing analysis 
of four clusters find 
$M/L_{R}\;\simeq 190\;h_{50}\; M_{\odot}/L_{\odot}$.
We seem to confirm a dichotomy in recent determinations of cluster 
$M/L$ ratios which in our case is due to X-ray masses 
systematically lower than virial masses.

Since galaxy clusters are the deepest potential wells in the Universe,
and they accumulate baryonic and dark matter from large volumes, it is
assumed that their $M/L$ ratio should represent 
the Universal value. Therefore,
in order to derive the value of $\Omega_{\rm m}$ we need to derive 
the universal luminosity density, $j$, by integrating the $R$-band
luminosity function of field galaxies. 
Although the functional form of the field galaxy LF is a Schechter function, 
the LF parameters vary from survey to survey (Brown et al.\ 2001). 
The only field-galaxy survey with the same observational set-up 
(filters, CCD detector, telescope) as LOCOS,
is the Century survey (CS; Geller et al.\ 1997).
The resulting Schechter LF (Brown et al.\ 2001) has parameters~
$\alpha \simeq -1.07$, 
$\phi_{*}=2.0\times\,10^{-3}\,h_{50}^{3}\,{\rm Mpc}^{-3}$ and 
$L_{R,*}=4.64\times\,10^{10}\,h_{50}^{-2} L_{\odot}$
which then gives 
$j_{R, \rm CS} \simeq 9.26\times\,10^{7}\;h_{50}\;L_{\odot},$ in good 
agreement with the value derived from 2dFGRS (Norberg et al.\ 2002).
Then the universal $M/L$ ratio is given by~
$\rho_{\circ}/j=\Omega_{\rm m}\,\rho_{\rm cr}/j 
 \simeq 760\,\Omega_{\rm m}\,h_{50}\;M_{\odot}/L_{\odot},$
where the critical density is
$\rho_{\rm cr}=7.016\times\,10^{10}\;h_{50}^{2}\;M_{\odot}\;{\rm Mpc}^{-3}$. 
As a further comparison we note that the Sloan Digital Sky Survey (SDSS)
$r^{*}$ luminosity function, extrapolated to redshift $z=0$ and
based on $\sim 150000$ galaxies (Blanton et al.\ 2003), has 
$\alpha = -1.05$, $\phi_{*} = 1.86 \times 10^{-3} h_{50}^{3} \mbox{ Mpc$^{-3}$}$ and $ L_{r,*} = 
4.9 \times 10^{10} h_{50}^{-2} L_{\odot}$. Shifting to the Cousins-$R$ band 
used in LOCOS, according to $R-r^{*} = 0.22$ (Fukugita et al.\ 1996), 
we obtain a larger value of the luminosity density:
$j_{R, \rm SDSS} \simeq 1.13 \times 10^{8} \; h_{50} \;
L_{\odot}$, which results in a universal $M/L \sim 620 \; \Omega_{\rm m} 
\; h_{50} \; M_{\odot}/L_{\odot}$.

It is now straightforward to obtain the value of $\Omega_{\rm m}$,
which for the case of the virial cluster masses is~
$\Omega_{\rm m} \simeq 0.23 ~{\rm or}~ 0.28\,(\pm 0.08),$
for the CS and SDSS luminosity functions, respectively.
The corresponding values for the X-ray based cluster masses are~
$\Omega_{\rm m} \simeq 0.13 ~{\rm or}~ 0.16\,(\pm 0.02).$
Only the former values are consistent with the {\em concordance} 
Cosmological model.

\section{Summary}
We used the virial theorem to estimate the total gravitating mass 
for a sample of 459 Abell/ACO clusters with robust velocity dispersions.
Their $R$-band luminosities were estimated from a relation between 
Abell galaxy counts and luminosity, based on 46 Abell clusters 
with accurate  $R$-band CCD photometry. The resulting median $M/L$ ratio
of the subsample analysed is $\simeq 178 \; h_{50} M_{\odot}/L_{\odot}$, 
if virial masses are used, and 
$M/L_{R} \simeq 100 \; h_{50} M_{\odot}/L_{\odot}$, 
if the X-ray based cluster masses are used.
Comparing with the universal $M/L$ value, we obtain
a mass content of the Universe of $\Omega_{\rm m} \sim
0.26$ and $\sim 0.15$, respectively.

\acknowledgments 
HA thanks CONACyT Mexico (grant 40094-F) and the meeting organizers for support.
MP is supported by grant CONACyT-2002-C01-39679, and OLC by CONACyT 
Young Researcher Grant J32098-E.

\end{document}